\documentstyle[prd,aps,preprint]{revtex}
\tightenlines
\input epsf.tex
\def\DESepsf(#1 width #2){\epsfxsize=#2 \epsfbox{#1}}

\begin{document}

\draft
\preprint{\vbox{
\hbox{OSU-HEP-99-01}\hbox{CTP-TAMU-13-99}
\hbox{UMD-PP-99-103}
}}
\title{Up--Down Unification, Neutrino Masses, and Rare Lepton Decays}

\author{ K. S. Babu$^1$, B. Dutta$^2$ and R. N. Mohapatra$^3$ }

\address{$^1$Department of Physics, Oklahoma State University,  Stillwater, OK
74078\\
$^2$Department of Physics, Texas A \& M University, College Station, TX  77843\\
$^3$Department of Physics, University of Maryland, College Park, MD 20742.}
\date{April, 1999}
\maketitle

\begin{abstract}

In a recent paper, we showed that tree level up-down
unification of  fermion Yukawa couplings is a natural consequence of a large class of
supersymmetric models.  They can lead to viable quark masses and mixings
for moderately large values of $\tan\beta$ with interesting and testable predictions for CP
violation in the hadronic sector. In this  letter, we extend our discussion
to the leptonic sector focusing on one particular class of these models,
the supersymmetric left--right
model with the seesaw mechanism for  neutrino masses. We show that fitting the
solar and the atmospheric neutrino data considerably restricts the
Majorana--Yukawa couplings of the leptons in this model and leads to predictions
for the decay  $\tau\rightarrow \mu +\gamma$, which is found
to be accessible to the next generation of rare decay
searches. We also show that the resulting parameter space of the model is
consistent with the requirements of generating adequate baryon asymmetry
through lepton--number violating decays of the right--handed neutrino.
\end{abstract}

\vskip1.0in

The discovery of neutrino mass has added a new impetus to the search for new
physics not only beyond the standard model but also beyond the minimal
supersymmetric extension of the standard model (MSSM). One  candidate for such a
theory is the supersymmetric version\cite{susylr,pheno} of the left--right
symmetric model\cite{lrs}, which provides a natural embedding of the seesaw
mechanism for small neutrino masses. There are two ways of realizing the seesaw
mechanism in these models: One by using the $B-L=2$ iso--triplet Higgs fields
in which case the seesaw arises from purely renormalizable interactions;
another using $B-L=1$
iso--doublet Higgses to get seesaw out of nonrenormalizable interactions.
The first alternative has the merit that it
results in automatic $R$--parity as a consequence of
gauge invariance.  The
second alternative has the apparent advantage that known compactification
schemes of heterotic string models can lead to this class of models at low
energies. The disadvantage is that one needs extra symmetries to maintain the
known degree of baryon and lepton number conservation (equivalently $R$--parity).

In a recent paper\cite{bdm}, we have shown that in both these versions of the
SUSY left--right models (as well as in their generalizations to $SO(10)$),
if the theory below the right handed
$SU(2)_R$ scale is assumed to be given by the MSSM, then one ends up  with a
strong tree level relation between the up and down quark and lepton Yukawa
coupling matrices, i.e.,
\begin{eqnarray} Y_u~=Y_d; ~~~~~~~~~~~~~~~Y_{\ell}~=~Y_{\nu^D}.
\end{eqnarray} We call this up-down unification\cite{posp}. These relations
follow for arbitrary Higgs structure in the first class of models (i.e. the ones
with $B-L=2$ HIggs fields) whereas in the second class of models they hold only in the
minimal version with one bi--doublet field. Furthermore, in the second class of
models, there is an overall constant of proportionality instead of equality. In
any case, as is clear, these relations lead to vanishing of the quark mixing
angles (the CKM angles) at the tree level.  They also lead to the
proportionality predictions for
the quark masses ($m_u:m_c:m_t = m_d:m_s:m_b)$
which do not agree with observations. As was demonstrated in
Ref.\cite{bdm}, once the one loop corrections coming from gluino
exchanges (which are known to be significant in such models, especially
when $\tan\beta$ is large \cite{raby}) are included, the
observed mixing and masses for the quarks can be obtained for a particular
choice of flavor mixing in the squark sector.  It was shown in Ref. \cite{bdm}
that the flavor changing neutral currents that are associated with such squark
mixings are consistent with experimental limits.
Further, it was pointed out  that these
models lead to specific predictions for the CP-violating parameters
$\epsilon'/\epsilon$ and electric dipole moment of the neutron which are
different from the conventional MSSM so that they can be used to test these
models.  Additional tests of these models are provided by branching ratios
of Higgs decay into fermion pairs, and deviation from the standard model
predictions in rare processes such as $B_d-\overline{B}_d$ mixing.

It is the goal of this paper to explore the implications of up--down unification
in the lepton sector. The interest in this problem stems from the fact that
up--down unification reduces the number of parameters in  the seesaw formula for the
neutrino masses without the assumption of grand unification. For instance, in a
generic seesaw model for three generations, there are twelve (complex) parameters
describing  the complete seesaw matrix for the light neutrinos (three from the
heavy Majorana neutrino sector and nine from the Dirac mass matrix). If one
assumes that the neutrino Dirac mass is symmetric (as can happen in some GUT
models or left--right models), we are still left with 9 parameters. On the other
hand, in models with up--down unification that we discuss in this paper there
are only six parameters. This is because the Dirac neutrino masses in a particular
basis is diagonal and completely given by the charged lepton masses. This
narrows down the possible parameter space in which a joint fit to the solar and
atmospheric neutrino data can be obtained. We find that the Majorana
Yukawa couplings that fit data exhibit a hierarchical pattern similar to that in the
charged fermion sector of the standard model.  Such a hierarchical structure in
the heavy Majorana neutrino mass matrix goes well with the idea of baryogenesis
through lepton asymmetry induced by the decay of the lightest right--handed
neutrino ($N_e$).
We use the neutrino fits to make
predictions for the rare decay $\tau\rightarrow \mu +
\gamma$  as a function of the seesaw scale (denoted by $v_R$).  Such rare
decays have a supersymmetric origin:  In running the soft sfermion masses
from the scale of gravity ($M_{\rm Planck}$ or $M_{\rm string}$)
to $v_R$, the right--handed neutrinos are ``active"
through their Majorana Yukawa couplings, which induce flavor mixing in the
slepton sector.  We
find that there exists a range of $v_R$ below the GUT scale for
which the predicted $\tau\rightarrow \mu +\gamma$ branching ratio is within the
reach of the planned experiments.

Let us start by giving a brief derivation of the up-down unification relation in
the SUSY left--right models. As is well known, the electroweak gauge group for this model is
$SU(2)_L\times SU(2)_R\times U(1)_{B-L}$ with the standard assignment where
$Q,Q^c$ denote left--handed and right--handed quark doublets, $\Phi_i$ denote the
(2,2,0) Higgs bi--doublets, $\Delta$ and $\Delta^c$ are the left-- and
right--handed Higgs triplets and $\bar{\Delta}$ and $\bar{\Delta^c}$ fields, their
conjugates. Let us write down the gauge invariant  part of the matter
superpotential involving these fields:
\begin{eqnarray}
W & = &
{\bf h}^{(i)}_q Q^T \tau_2 \Phi_i \tau_2 Q^c +
{\bf h}^{(i)}_l L^T \tau_2 \Phi_i \tau_2 L^c
\nonumber\\
  & +  & i ( {\bf f} L^T \tau_2 \Delta L + {\bf f}_c
{L^c}^T \tau_2 \Delta^c L^c).
\end{eqnarray}
Left--right symmetry implies ${\bf h}^{(i)}_{q,l} = ({\bf h}^{(i)}_{q,l})^\dagger$ and
${\bf f} = {\bf f_c}$.
It is clear from the above equation that if there is only a
single bidoublet field $\Phi (2,2,0)$, then up--down unification will follow. As has
been shown in Ref.\cite{bdm}, this result holds even when more than one
bi--doublet is included in the theory. The result is more general if $B-L=2$
Higgs fields are employed for $SU(2)_R$ gauge symmetry breaking.
The same result
holds if the right handed symmetry is broken by
$B-L=1$ doublets as well, but if there is only one Higgs bi--doublet. What happens in this
case is that at the $v_R$ scale, the two doublets ($\phi_{u,d}$) from the
bi--doublet $\Phi$ and the two
left--handed doublet superfields transforming as $(2,1,\pm1)$ under
$SU(2)_L \times SU(2)_R \times U(1)_{B-L}$ (call them $\chi$ and $\bar{\chi}$) form a
$2\times 2$ mass matrix. Once we assume that appropriate fine tuning has been
done to leave the MSSM doublets (denoted by $H_u$ and $H_d$)
light below the $v_R$ scale, their generic form
becomes $H_u\equiv \cos \theta \phi_u + \sin\theta \bar{\chi}_u$ and
$H_d\equiv \cos\alpha \phi_d +\sin\alpha \chi_d$. In this case, we obtain  the
modified up-down unification formula $Y_u = \gamma Y_d$ and
$Y_{\ell}=\gamma Y_{\nu^D}$, where $\gamma=\cos\theta/\cos\alpha$
is a proportionality constant.

Let us now study the implications of up-down unification in the lepton sector.
We pass to a basis where the charged lepton mass matrix is diagonal. In this
basis one gets:
\begin{eqnarray} M_{\nu^D}= \tan\beta ~M_{\ell}
\end{eqnarray} for the renormalizable seesaw case (triplet case) and the same
relation with an extra constant multiplier in the doublet case. Here, $M_{\ell}=
Diag (m_e, m_{\mu}, m_{\tau})$. The light Majorana neutrino mass matrix is then
given by:
\begin{eqnarray} M_{\nu}=-\frac{\tan^2\beta}{v_R} M_{\ell} {\bf f}^{-1} M_{\ell}
\end{eqnarray}
where ${\bf f}$ is the right--handed neutrino
Majorana Yukawa coupling defined in Eq. (1). In the case of using doublet Higgs fields,
$\bf f$ happens to be the coupling associated with the
non-renormalizable interaction of the right handed neutrinos i.e.
$f_{ij}(L^c_i\chi^cL^c_j\chi^c)/M$ and analogous discussions will follow.
Eqs. (2)-(4) parameterize the neutrino masses
and mixings in terms of six parameters. All parameters are complex in general,
but for simplicity of presentation we shall assume them to be real.

To study the neutrino masses and mixings in this model, let us start with
a specific parameterization of the $f$ matrix as follows:
\begin{eqnarray}  {\bf f}&=&\left(\begin{array}{ccc}
\lambda & \delta & 0 \\
\delta & \epsilon^2 & b\epsilon \\ 0 & b\epsilon & 1 \end{array} \right).
\end{eqnarray}
This is the most general form of a symmetric $3 \times 3$ matrix, except for
the $0$ in the $(1,3)$ entry.
This choice has been motivated by considerations of
baryogenesis, discussed later in this article, which requires
$f_{13}$ to be very small.  Eq. (5) leads, after see--saw diagonalization, to a
Majorana neutrino mass matrix given by:
\begin{eqnarray} M_{\nu}=\frac{tan^2\beta}{v_R
g(\epsilon,\delta,\lambda,b)}\left(\begin{array}{ccc}
-\frac{\epsilon^2}{\lambda}(1-b^2)m^2_e & \frac{\delta}{\lambda} m_em_{\mu} &
-\frac{b\epsilon\delta}{\lambda}m_em_{\tau} \\
\frac{\delta}{\lambda}m_em_{\mu} & -m^2_{\mu} & m_{\mu}m_{\tau} b\epsilon
\\ -\frac{b\epsilon\delta}{\lambda}m_em_{\tau} & b\epsilon m_{\mu}m_{\tau} &
-m^2_{\tau}(\epsilon^2 - \frac{\delta^2}{\lambda}) \end{array}\right)~.
\end{eqnarray} Here $g(\epsilon, \delta,
\lambda,b)=\frac{\delta^2}{\lambda}-\epsilon^2(1-b^2)$. Note that the neutrino
masses and mixings depend on four independent parameters which are completely
fixed by observations, thereby fixing  all parameters of the lepton sector of
the theory. Also note the suppression factors arising from
the charged lepton masses in Eq. (6).

Recently, the Super-Kamiokande collaboration \cite{kam} has established
a deficit in the measured flux of muon neutrinos from the atmosphere.  This
observation can
be interpreted as compelling evidence in favor of neutrino masses and
oscillations. The most likely scenario is oscillation of $\nu_\mu \rightarrow
\nu_\tau$.  The mass--splitting is inferred to be: $\Delta m^2_{\mu\tau}\simeq
(10^{-3}-10^{-2})$ eV$^2$ and the oscillation angle is in the range:
$\sin^22\theta_{\mu\tau}\simeq 0.8-1$. We
will assume parameters consistent with this interpretation in our analysis. The
long--standing solar
neutrino deficit can be explained by neutrino oscillation in matter
(MSW oscillation \cite{msw}) between $\nu_e$ and another species (most likely
$\nu_\mu$ if there are no extra species of neutrinos, as in our models).  Two possible
solutions exist: the small angle MSW and large angle MSW
regions.  We shall adopt the small angle solution
for which
$\Delta m^2_{e\mu}\simeq (0.3-1)\times 10^{-5}$ eV$^2$ and $2\times
10^{-3}\leq\sin^22\theta_{e\mu}
\leq2\times 10^{-2}$. To see how one obtains  solutions that fit both solar and
atmospheric neutrino data, let us first focus on the $\nu_{\mu}-\nu_{\tau}$
sector and assume that $\epsilon^2 \gg
\frac{\delta^2}{\lambda}$. In this case, it is clear that if $\epsilon
\simeq \frac{m_{\mu}}{m_{\tau}}$, then for $b\simeq 1$ we have maximal mixing
between the $\nu_{\mu}$ and $\nu_{\tau}$. Furthermore, if $b=1$, we have
$m_{\nu_{\mu}}=0$. Therefore if we take $b=1+\kappa$ with $\kappa\ll 1$, we get
\begin{eqnarray}
\frac{m_{\nu_{\mu}}}{m_{\nu_{\tau}}}\simeq \frac{\kappa}{2}
\end{eqnarray}
 and the $\nu_{\mu}-\nu_{\tau}$ mass difference sqared given by:
\begin{eqnarray}
\Delta m^2_{\mu \tau}\simeq m^2_{\nu_{\tau}}\simeq
\frac{\tan^4\beta}{v^2_R}
\frac{(4b) m^4_{\tau}}{\kappa^2}.
\end{eqnarray} We thus see that $m_{\nu_{\mu}}\ll m_{\nu_{\tau}}$ and the
atmospheric neutrino data fixes $m_{\nu_{\tau}}$. To get solar neutrino via the
MSW  small angle solution, we require $\kappa\simeq 0.1$ or so. The formula for
$m_{\nu_{\tau}}$ then enables us to determine the value of $v_R$ and we find it
to be around $10^{15}$ GeV if we choose
$\tan\beta \simeq 40$, which seems required from considerations of the quark
mixings. Let us hasten to note that this is not a complete determination of the
seesaw scale -- since we have scaled the Majorana coupling $f_{33}=1$ to get this
value. There is an overall scaling freedom and we could scale all the $f_{ij}$'s
by a common factor that would change the $v_R$. For instance, if we scaled down
all the
$f_{ij}$'s by a factor of 100, we would get $v_R\simeq 10^{17}$ GeV which  is
near the GUT scale. If we limit the value of $v_R$ to stay below the string
scale ($\sim 4 \times 10^{17}$ GeV),
then we have a small range for $f_{33}$ between 1-0.01. Thus, although there is
no dynamical mechanism in the model that would independently fix the overall
scale of the $f_{ij}$'s, we do not have too much freedom in their range.

The solar neutrino puzzle has also a solution in this specific parameterization.
To see this, note that one easily finds $m_{\nu_e}\ll m_{\nu_{\mu}}$ and we have
adjusted the value of $\kappa$ to get the right mass difference squared between
$\nu_\mu$ and $\nu_e$. The mixing has a suppression factor of $m_e/m_{\mu}$ but
is multiplied by the ratio
$\delta/\lambda$ so that we can adjust those parameters to get the required mixing
angle needed in the small angle MSW solution. We do not mean to imply that this is
a unique solution -- but rather we present this example to illustrate the general
nature of the detailed numerical fits using the parameterization for the $\bf f$
matrix. Below we give the values for the  different elements of the $\bf f$
matrix that lead to acceptable solutions to both the solar and the atmospheric
neutrino data.

With $\tan\beta=41$ and $v_R=10^{14.9}$ GeV we choose
\begin{eqnarray} {\bf f}&=&\left(\matrix{
  -7.9\times10^{-6}                & 7\times 10^{-5}   &0     \cr
  7\times 10^{-5}            &-2.5\times 10^{-3}       &4.8\times 10^{-2}
\cr
               0                &4.8\times 10^{-2}        &-1   }\right);
\end{eqnarray}
which gives rise to neutrino masses: $\left(\matrix{
 4.3\times 10^{-5},                & 2.9\times 10^{-3},   & 6.1\times
10^{-2}       }\right)$ eV  and the leptonic mixing matrix:
\begin{eqnarray} U &=&\left(\matrix{
 -0.99                & 4.3\times 10^{-2}   & -3.8\times 10^{-4}    \cr
  2.8\times 10^{-2}            &0.64      &-0.77 \cr
              -3.8\times 10^{-2}               &-0.77       &-0.64
}\right);
\end{eqnarray}
In our notation, $U_{12}$ is the mixing angle relevant for solar neutrinos.

We have also obtained consistent fits including non--zero $f_{13}$.  For example,
choosing
\begin{eqnarray} {\bf f}&=&\left(\matrix{
  -1.1\times10^{-5}                & -1.0\times 10^{-4}   &3.8\times 10^{-3}
\cr
  -1.0\times 10^{-4}            &-1.7\times 10^{-3}       &4.8\times 10^{-2} \cr
     3.8\times 10^{-3}          &4.8\times 10^{-2}        &-1.56   }\right).
\end{eqnarray}
gives rise to following neutrino masses at
$v_R=10^{14.9}$ GeV (for $\tan\beta = 41$):
$(9.8\times 10^{-6},\, 3.0\times 10^{-3},\,  6.4\times
10^{-2})$ eV.  The leptonic mixing matrix is given by:
\begin{eqnarray} U = &&\left(\matrix{
 -0.99                & -4.46\times 10^{-2}   & 8.9\times 10^{-2}    \cr
  9.4\times 10^{-2}            &-0.70      &0.71 \cr
              3.2\times 10^{-2}               &0.71       &0.71   }\right).
\end{eqnarray}
There are small corrections to these values coming from the
renormalization group extrapolation from the $v_R$ scale to the weak scale
\cite{babu}.

Let us now comment on the baryogenesis constraint on the matrix elements of
$\bf f$. In the
picture where baryogenesis results from the leptogenesis caused by the decay of
heavy Majorana neutrinos in the very hot stage of the universe\cite{fuku}, one
needs these neutrinos to be out of equilibrium. It is
well known
that in these scenarios, only the lightest of the three right handed neutrinos
(the $N_e$) is responsible for leptogenesis. In the presence of $f_{13}$, its
decay has a contribution coming from the $\tau + H_u^+ $ final state and is
given by
\begin{eqnarray}
\Gamma (N_e\rightarrow \tau\nu_{\tau})\simeq \frac{(f_{13}m_{\tau}
\tan\beta)^2}{12\pi v^2} M_{N_e}~.
\end{eqnarray}
Furthermore, the mass of the $N_e$ must be less than the re--heating
temperature of the universe after inflation, which is required by
gravitino abundance constraints to be less than $10^{9}$ GeV. Putting this
value in the out of equilibrium condition
$\Gamma({N_e})\leq H$ requires that $f_{13} \leq 10^{-3}$. Thus the $f_{13}$ element
must indeed be very small and this justifies our choice of zero for this
parameter.  With the form of the heavy Majorana neutrino matrix fixed from
low energy neutrino oscillation data, we can examine if the required
lepton asymmetry can be induced in the decay of $N_e$ \cite{fuku}.  We
find that adequate lepton asymmetry, as large as $10^{-6}$, can be easily
induced with the choice of parameters given in Eqs. (9), (11).  The actual
baryon asymmetry will be quite sensitive to the dilution factor.  Given
that a lepton asymmetry of order $10^{-6}$ can be realized in our model,
some dilution arising from $N_e$ not totally being out of equilibrium will
be allowed.  This is the reason that in the second example (Eq. (11)) we
allowed a small $f_{13}$ entry which is marginally inconsistent with the
bound in Eq. (13).

\vskip0.1cm
\noindent {\bf Predictions for Rare $\tau$ and $\mu$ decays}

\vskip0.2cm

Having obtained the various elements of the matrix $\bf f$, let us turn to the
phenomenological implication of this result in the leptonic sector. We
shall assume that the slepton masses are universal at the Planck or GUT scale. Now due
to the presence of off--diagonal elements in  $\bf f$, the renormalization
group evolution will induce mixings between the second and the third
generation sleptons (i.e. terms such as
$\tilde{\mu}^{\dagger}\tilde{\tau}$) as well as between the first and the second generation
sleptons.  Such mixings arise because between the Planck/GUT scale (where
universality holds) and $v_R$, the Majorana Yukawa couplings will contribute to
the evolution of the soft parameters.
The $\tilde{\mu}-\tilde{\tau}$ mixing will depend on the product
$f_{33}f_{23}$. Similarly, there will be mixing between $\tilde{e}$ and
$\tilde{\mu}$ proportional to $f_{11}f_{12}$ as well as $f_{13}f_{23}$. These off
diagonal slepton mass terms are well known\cite{hall} to induce rare lepton
decays  such as $\tau\rightarrow \mu +\gamma$ and $\mu\rightarrow e +\gamma$.

There are two possible diagrams in lepton flavor violating  decay modes in this
model: (i) the chargino ($\chi^{\pm}$)  mediated (involving neutral sleptons) and
(ii) the neutralino ($\chi^{0}$) mediated (involving charged sleptons). The
chargino diagram is comparable with the neutralino diagram. In the  chargino
diagram, we have contributions only from the left--handed sneutrinos. This is
because
 only the left--handed sneutrinos survive down to the weak scale. On the
other hand the left as well the right--handed charged sleptons contribute to the
neutralino diagram. We will see that the masses of both the helicities develop
flavor violating soft terms through RGE due to the presence of off diagonal
elements in $\bf f$.

$l_j\rightarrow\l_i+\gamma$ has the following effective lagrangian
($m_{l_j}>m_{l_i}$):
\begin{eqnarray} L_{eff}={m_{l_j}\over 2}(A_1\bar
l_i\sigma^{\mu\nu}P_R\l_jF_{\mu\nu}+ A_2\bar
l_i\sigma^{\mu\nu}P_L\l_jF_{\mu\nu})+h.c.\end{eqnarray} where
$A_1=A1_{\chi^{\pm}}+A1_{\chi^{0}}$ and $A_2=A2_{\chi^{\pm}}+A2_{\chi^{0}}$.

The decay width is
given by:
\begin{eqnarray}
\Gamma(l_j\rightarrow l_i\gamma)=
{m_{l_j}^5\over(16\pi)}(\left|A_1\right|^2+\left|A_2\right|^2).\end{eqnarray}

$A1_{\chi^{\pm}}$ and $A1_{\chi^{\pm}}$ are given by:
\begin{eqnarray}A1_{\chi^{\pm}}&=&\alpha_W {{\sqrt\alpha}\over{2 {\sqrt \pi}}}
         \sum_{m=1}^2\sum_{k=1}^3 {1\over m^2_{\tilde{\nu_k}}}\times\\\nonumber
&&
         \left\{G_{\nu L}^{m,k,j} G_{\nu
L}^{*m,k,i}F_2(x_{{\chi^{\pm}_m}\tilde\nu_k})-
          H_{\nu L}^{m,k,j}G_{\nu L}^{* m,k,i} {M_{\chi^{\pm}_m}\over m_j}
          F_3(x_{{\chi^{\pm}_m}\tilde\nu_k})\right\},
\end{eqnarray}

\begin{eqnarray}A2_{\chi^{\pm}}&=&\alpha_W {{\sqrt\alpha}\over(2 {\sqrt \pi})}
         \sum_{m=1}^2\sum_{k=1}^3 {1\over m^2_{\tilde{\nu_k}}}\times
         \left\{H_{\nu L}^{m,k,j} H_{\nu L}^{*m,k,i}
     F_2(x_{{\chi^{\pm}_m}\tilde\nu_k})\right\},\end{eqnarray}  where
the convention $x_{ab}={m^2_a}/{m^2_b}$ has been adopted and $G_{\nu
L}^{m,k,j}=vC^{* m,1} \Gamma^{k,i}_{\nu L}$
$H_{\nu L}^{m,k,i}=uC^{ m,1}\left(\Gamma_{\nu L}{\hat U_l}\right)^{k,i}$.
$\Gamma^{k,i}_{\nu L}$ are the sneutrino rotation matrices and the
$uC$, $vC$ are the chargino rotation matrices and
${\hat U_l}=diag(m_e,m_\mu,m_\tau)/({\sqrt 2} M_W cos\beta)$. $M_{\chi^\pm_m}$
are the chargino masses and $m^2_{\tilde\nu_k}$ are the sneutrino masses.
$F_2(x)=1/(12((x-1)^4)(2 x^3+3 x^2-6x+1-6x^2{\rm ln} x)$ and
$F_3(x)=1/(2((x-1)^3)(x^2-4x+3+2 {\rm ln} x)$.

 The
expressions for the neutralino contributions are \cite{ddk}.
\begin{eqnarray} A1_{\chi^0}&=&-{{\alpha \sqrt{\alpha}}\over {2\cos^2 \theta_W
\sqrt{\pi}}}\sum_{m=1}^4\sum_{k=1}^6
\frac1{M_{\tilde{l_k}}^2}\times\\\nonumber && \left\{\left(\sqrt
2G^{mkj}_{0lL}\right) \left(\sqrt 2 G^{*jki}_{0lL}\right) F_2\left(x_
{\tilde{\chi_{m}^0}\tilde{l_k}}\right) -\right. \\\nonumber && \left.\left(\sqrt
2G^{mk\j}_{0lR}-H^{mk\j}_{0lL}\right) \left(\sqrt 2
G^{*mki}_{0lL}\right)\frac{m_{\tilde{\chi^0_m}}}{m_j}
F_4\left(x_{\tilde{\chi_{m}^0}\tilde{l_k}}\right) \right\},\\\nonumber
A2_{\chi^0}&=&-{{\alpha \sqrt{\alpha}}\over {2\cos^2 \theta_W
\sqrt{\pi}}}\sum_{m=1}^4\sum_{k=1}^6
\frac1{M_{\tilde{l_k}}^2}\times\\\nonumber && \left\{\left(\sqrt
2G^{mkj}_{0lR}\right)\left(\sqrt 2 G^{*mki}_{0lR}\right) F_2\left(x_
{\tilde{\chi_{m}^0}\tilde{l_k}}\right) -\right. \\\nonumber && \left.\left(\sqrt
2G^{mkj}_{0lL}+H^{mkj}_{0lR}\right) \left(\sqrt 2
G^{*jki}_{0lR}\right)\frac{m_{\tilde{\chi^0_m}}}{m_j}
F_4\left(x_{\tilde{\chi_{m}^0}\tilde{l_k}}\right) \right\}, \end{eqnarray}
where
\begin{eqnarray} G^{mkh}_{0lL}&=&-1/2[Z_{m1}+\cot\theta_W
Z^*_{m2}]\Gamma^{kh}_{lL}\\\nonumber
G^{mkh}_{0lR}&=&-Z_{m1}\Gamma^{kh}_{lR}\\\nonumber
H^{mkh}_{0lL}&=&Z_{m3}\left(\Gamma_{lL}\hat{U}_l\right)^{kh}\\\nonumber
H^{mki}_{0lR}&=&Z_{m3}\left(\Gamma_{lR}\hat{U}_l\right)^{kh}\, ,\\\nonumber
\end{eqnarray} where
$Z$ is the
$4\times 4$ neutralino mixing matrix in the $\left(\tilde{B},
\tilde{W_3},\tilde{H^0_1},\tilde{H^0_2}\right)$ basis,
 $F_4(x)=1/(2((x-1)^3)(x^2-1-2x{\rm ln} x)$ and $\tilde{l}_{L,R}
=\Gamma^{\dagger}_{lL,R} \tilde{l}$. $M_{\tilde{l_k}}$
are the slepton mass eigenstates. $\tilde{l}_{L,R}$ is $6\times 6$ slepton
mass matrix $\tilde{l}_{L,R}$ at the weak scale  and can be written  in
the $3\times 3 $ forms having the submatrices $M_{LL}$,$M_{LR}$ and $M_{RR}$ as
follows:\begin{eqnarray}
\left(\begin{array}{cc} M_{lL}M_{lL}^{\dag} & M^2_{lLR} \\ M^{2\dagger}_{lLR} &
M_{lR}^{\dag}M_{lR}\end{array}\right) \label{slmass}\, , \end{eqnarray}
where \begin{eqnarray}\left(M_{lL}M_{lL}^{\dag}\right)_{ij}&=&
\left({\bf m^2_{LL}}\right)_{ij}+
M^2_Z\cos2\beta\left({1\over 2}-\sin^2\theta_W\right)\delta_{ij}\, ,\\\nonumber
\left(M^2_{lLR}\right)_{ij}&=&\left(U_{l_A} v_d+U_l\,\mu\, tan\beta\, v_d\right)_{ij}\, ,\\\nonumber \left(
M_{lR}^{\dag}M_{lR}\right)_{ij}
&=&\left({\bf m^2_{eR}}\right)_{ij}+M^2_Z\cos2\beta\left(-1\sin^2\theta_W\right)\delta_{ij}\,
.\\\nonumber
\end{eqnarray}Here $U_l$ is the lepton Yukawa matrix, $U_{l_A}=A_lU_l$  and
the $A$'s are the trilinear couplings present in the potential. We determine the
soft masses and the trilinear couplings at the weak scale by using the RGE,
which will be discussed next.

Before we calculate the branching ratios of $\tau\rightarrow\mu+\gamma$, let us
first discuss   the parameter space of this model. We assume the model
originates at the string scale ($M_{\rm string}\sim 10^{17}$ GeV) and is left-right
symmetric.  The trilinear terms involving the $\Delta$ fields, bi--doublets and the
matter fields in the superpotential are shown in Eq. (2). We observe that the
$\Delta$ and $\Delta^c$ have couplings given by $\bf f$ and $\bf f_c$ to the
leptons (${\bf f_c} = {\bf f}$ due to left--right symmetry).
This  superpotential is valid
upto the scale $v_R$. At this scale all the
$\Delta$ fields  and the right--handed
neutrinos pick up mass.  The doubly charged components of $\Delta$ and $\Delta_c$
fields do not acquire masses of order $v_R$, their masses arise through
non--renormalizable operators and are of order $v_R^2/M_{\rm Planck}$.  This
situation is different if we employ $SU(2)$ singlet doubly charged Higgs fields,
which may be desirable from the point of view of inducing electroweak symmetry
breaking (see comments later).
The RGE of the soft supersymmetry breaking slepton
masses are given by:
\begin{eqnarray} {d{\bf m}^2_{LL}\over dt}&=&{2\over {16 \pi^2}}[-4\pi(3/2
\alpha_{B-L} M_{B-L}^2+3 \alpha_{L}
M_{L}^2)\\\nonumber&+&{1\over2}((U_lU_l^{\dag}+{\bf f}\,{\bf f}^{\dag}){\bf
m}^2_{LL}+{\bf m}^2_{LL}(U_lU_l^{\dag}+{\bf f}\,{\bf f}^{\dag})+ 2(U_l{\bf
m}^2_{LL}U_l^{\dag}) \\\nonumber&+& 2({\bf f}{\bf m}^2_{LL}{\bf f}^{\dag} +
m^2_{\phi}U_lU_l^{\dag}+m^2_{\Delta}{\bf f}\,{\bf
f}^{\dag}+U_{l_A}U_{l_A}^{\dag}+U_{f_A}U_{f_A}^{\dag}))].
\end{eqnarray}  The same RGE also holds for the right--handed sleptons because of
left--right symmetry. It is easy to see that the terms like ${\bf f}\,{\bf
f}^{\dag}{\bf m}^2_{LL}$, $U_{f_A}U_{f_A}^{\dag}$ ($U_{f_A}=A_f{\bf f}$) introduce  flavor
violations in the soft terms for the left and the right handed sleptons. However,
the amount of flavor violations that is introduced through
$U_{f_A}U_{f_A}^{\dag}$ are rather small.

The  RGE's for the other soft terms, Yukawa couplings and  the trilinear terms
are as follows:
\begin{eqnarray} {d{\bf m}^2_{\phi}\over dt}&=&{2\over {16 \pi^2}}[-4\pi(3
\alpha_{R} M_{R}^2+3 \alpha_{L} M_{L}^2)\\\nonumber&+&3 \lambda_t^2 (2
m^2_{QL}+m^2_{\phi}+A_{Q}^2)\\\nonumber&+&2 Tr(m^2_{\phi}U_lU_l^{\dag}+U_l{\bf
m}^2_{LL}U_l^{\dag})],
\end{eqnarray}
\begin{eqnarray} {d{\bf m}^2_{\Delta}\over dt}&=&{2\over {16 \pi^2}}[-4\pi(4
\alpha_{L} M_{L}^2+4 \alpha_{B-L} M_{B-L}^2)\\\nonumber&+&2
Tr(m^2_{\Delta}U_lU_l^{\dag}+U_l{\bf m}^2_{LL}U_l^{\dag}+U_{f_A}U{f_A}^{\dag})],
\end{eqnarray}
\begin{eqnarray} {d U_{l_A}\over dt}&=&{1\over {16 \pi^2}}[-4\pi(3 \alpha_{R} +3
\alpha_{L} +{3\over 2}\alpha_{B-L} )U_{l_A}\\\nonumber&+&8\pi(3 \alpha_{R}
M_{R}+3 \alpha_{L} M_{L}+{3\over 2}\alpha_{B-L} M_{B-L})U_{L}\\\nonumber&+& 8
U_{l_A}U_l^{\dag}U_l +
    4 U_lU_l^{\dag} U_{l_A}+
    4 {\bf f}\,{\bf f}^{\dag}U_{l_A}+
    2 U_{f_A}{U_l}^{\dag}{U_l}\\\nonumber&+&
    2 Tr(U_{l_A}
    U_l^{\dag})U_l+
    Tr(U_l U_l^{\dag})U_{l_A} +
    6 A_{t}\lambda_t^2 U_l +
    3 \lambda_t^2 U_{l_A}],
\end{eqnarray}

\begin{eqnarray} {d U_{f_A}\over dt}&=&{1\over {16 \pi^2}}[-4\pi(7 \alpha_{R}
+{9\over 2}\alpha_{B-L})U_{f_A}+8\pi(7 \alpha_{R}M_{R} +{9\over
2}\alpha_{B-L}M_{B-L} ){\bf f}\\\nonumber&+&
    8 U_{l_A}U_l^{\dag}{\bf f} +
        4 U_lU_l^{\dag} U_{f_A}+
    4 U_{f_A}{\bf f}^{\dag}{\bf f}+
    2 {\bf f}\,{\bf f}^{\dag}U_{f_A}\\\nonumber&+&2 Tr(U_{f_A}
    {\bf f}^{\dag}){\bf f}+
    Tr({\bf f}\,{\bf f}^{\dag})U_{f_A}],\end{eqnarray}

\begin{eqnarray} {d {\bf f}\over dt}&=&{1\over {16 \pi^2}} [-4 \pi (7 \alpha_R
    +{9\over2} \alpha_{B-L}) {\bf 1}
    +2 {\bf f}\,{\bf f}^{\dag} +
    4 U_l U_l^{\dag}
    + Tr({\bf f}\,{\bf f}^{\dag}){\bf f}],
\end{eqnarray}
\begin{eqnarray} {d U_l\over dt}&=&{1\over {16 \pi^2}} [-4 \pi (3 \alpha_R+3
\alpha_L
    +{3\over2} \alpha_{B-L}) {\bf 1}
    +2 {\bf f}\,{\bf f}^{\dag} +
    4 U_l.U_l^{\dag}+Tr(3 U_q{U_q}^{\dag}
    + {U_l}{U_l}^{\dag}){U_l}].
\end{eqnarray}
$U_q$ is the up quark (or the down quark) Yukawa matrix. Because of the
left--right symmetry $\alpha_R$, $\alpha_L$, $\bf f$ and $\bf f^c$ renormalize
identically.  $M_i$s are the gaugino masses.

The doubly charged component of the
$\Delta^c$ fields, which got leaked through,  picks up mass at the scale
$M_{\Delta}(\sim {{v_R}^2\over M_{\rm string}}$). The superpotential, between $v_R$
and $M_{\Delta}$, contains a new interaction term (in addition to the MSSM
interactions) given by ${\bf f^c}l^c\Delta^{c\pm\pm}l^c$, where $l^c$'s are the
charged leptons. The right--handed slepton masses are affected by this
interaction term and the  RGE is given by:
\begin{eqnarray} {d{\bf m}^2_{eR}\over dt}&=&{2\over {16 \pi^2}}[-4\pi(12/5
\alpha_{1} M_{1}^2)\\\nonumber&+&((U_lU_l^{\dag}+{1\over 2}{\bf f^c}\,{\bf
f^c}^{\dag}){\bf m}^2_{eR}+{\bf m}^2_{eR}(U_lU_l^{\dag}+{1\over 2}{\bf f^c}\,{\bf
f^c}^{\dag})+ 2(U_l{\bf m}^2_{eR}U_l^{\dag}) \\\nonumber&+& ({\bf f^c}{\bf
m}^2_{eR}{\bf f^c}^{\dag} + 2 m^2_{H_1}U_lU_l^{\dag}+m^2_{\Delta^{c\pm\pm}}{\bf
f^c}\,{\bf f^c}^{\dag}+2 U_{l_A}U_{l_A}^{\dag}+U_{f^c_A}U_{f^c_A}^{\dag}))].
\label{delc}
\end{eqnarray}

There is a further twist to the story of soft masses in our model. The
constraint
$\lambda_t=\lambda_b$ at the left--right scale renormalizes the  up type and the
down type Higgs identically. It is almost impossible to make the (mass)$^2$ of
one of the Higgses to be positive and the other negative at the weak scale,
unless there exists some splitting of the up and down type Higgs masses already at the
left--right scale. Since in this model both the up type and  down type Higgs are
coming from the same bi--doublet, there is no scope to introduce  non--universality
in their masses at the string scale. However, there exists some natural splitting
for these masses at the left--right scale arising from  the D term. These D
terms  get generated due to the breaking of
$SU(2)_R\times U(1)_{B-L}\rightarrow U(1)_Y$ and depend on the mass difference of
the $\bar\Delta^c $ and the
$\Delta^c$ fields at the left--right scale. For example, the up type Higgs
mass--squared gets a
correction:
$-{{(m^2_{\Delta^c}-m^2_{\bar\Delta^c})}\over{4(1+ {3\over
2}(\alpha_{B-L}/\alpha_R))}}$ and the down type Higgs gets a negative of the
above correction. If the masses of $\Delta^c$ and $\bar{\Delta}^c$
are the same at the string scale,  $m^2_{\Delta^c}$ will be smaller than
$m^2_{{\bar\Delta^c}}$ due to renormalization, since $\Delta_c$ has
Majorana Yukawa couplings.   The mass--squared of the down type Higgs will then get a
negative contribution while the up type Higgs will get a positive contribution.
This goes against electroweak symmetry breaking requirement.
The  sign of this D term splitting is
fixed if universal boundary condition is assumed  at the string scale.
We can however
assume that the soft breaking masses for the $\Delta^c $ and the $\bar\Delta^c$
fields are not same (at the string scale)
and that these masses are are also different from the other soft
breaking masses.  A rationale for this assumption is discussed later.
We denote the departure from universality in the masses  by
$\delta_i$. We assume that the soft  mass squared term for the
$\Delta^c$ field at the string scale is given by
$m_0^2(1+\delta_1)$ and for the
$\bar \Delta^c$ field is given by $m_0^2(1+\delta_2)$. $m_0$ is the soft
breaking mass of the other scalars. With this assumption,  we find that the electroweak
symmetry can be broken in a large region of parameter space when we assume
$\delta_1$ to be positive and
$\delta_2$ to be negative. This non universality in the $\Delta$ masses will
also percolate down to other scalar masses at the $v_R$ scale through the  D
terms. The lighter slepton masses gets lowered and this  increases the
BR($\tau\rightarrow\mu+\gamma$). We will assume the gaugino masses to be
universal.

One possible source for the non--universality in the masses of $\Delta^c$
and $\overline{\Delta}^c$ at $v_R$ is that it might be dynamically
generated by Yukawa couplings.  As noted earlier, with the minimal
set of Higgs fields, the doubly charged components of $\Delta^c$
and $\overline{\Delta}^c$ acquire masses of order $v_R^2/M_{\rm Planck}$.
One way to give these members masses of order $v_R$ is to introduce
$(1,1,\pm 4)$  Higgs fields ($\eta, \overline{\eta}$).  Then the following
Yukawa couplings in the superpotential are allowed: $\lambda_1 \Delta^c
\Delta^c \eta + \lambda_2 \overline{\Delta}^c \overline{\Delta}^c \overline{\eta}$.
If $\lambda_2 \ll \lambda_1$, at $v_R$, the soft mass--squared of
$\overline{\Delta}^c$ will be smaller than that of $\Delta^c$.  The D term
splitting will then have the right sign to lower the mass of $H_u$ relative
to that of $H_d$, facilitating electroweak symmetry breaking.

For most part of our numerical analysis, we shall assume a spectrum without the
$\eta, \overline{\eta}$ fields.  We have verified that the major
consequence of including these fields can be taken care of by allowing
non--universal soft masses for the $\Delta^c$ fields at the Planck scale.
One example for this case is given in Fig. 6.

Using these masses as input at the string scale, we determine the spectrum of
this model at the weak scale. The procedure of computation goes as follows. We
first determine the Yukawa couplings and
$\bf f$ at the string scale using the experimental masses and mixing angles of
the quarks and leptons.  At the string scale we introduce the soft breaking masses
and run down all the masses and the couplings to the left--right scale using the
RGEs relevant to the left--right symmetry. From the
scale, at which the doubly charged fields get decoupled, down to
the weak scale, we use the MSSM RGEs(as given in ref.\cite{bbo} (we are also using
their sign convention of the $\mu$ term). In between the left--right and the doubly
charged field decoupling scale we use the MSSM RGEs coupled with the 
Eq.(\ref{delc}). At the weak scale we have the $ 6\times 6$
slepton mass matrices Eq.(\ref{slmass}). All the $3\times 3$ blocks are evaluated at
the weak scale for the charged slepton masses. For the neutral sleptons,
$M^2_{lLR}$ and the
$M_{lR}^{\dag}M_{lR}$ get decoupled at the left--right scale.

We also determine the other sparticle masses which are not much different from
what could be their masses in MSSM for a given value of
$m_0$, $m_{1/2}$, $\tan\beta$ and $A$ .

\begin{figure}[htb]
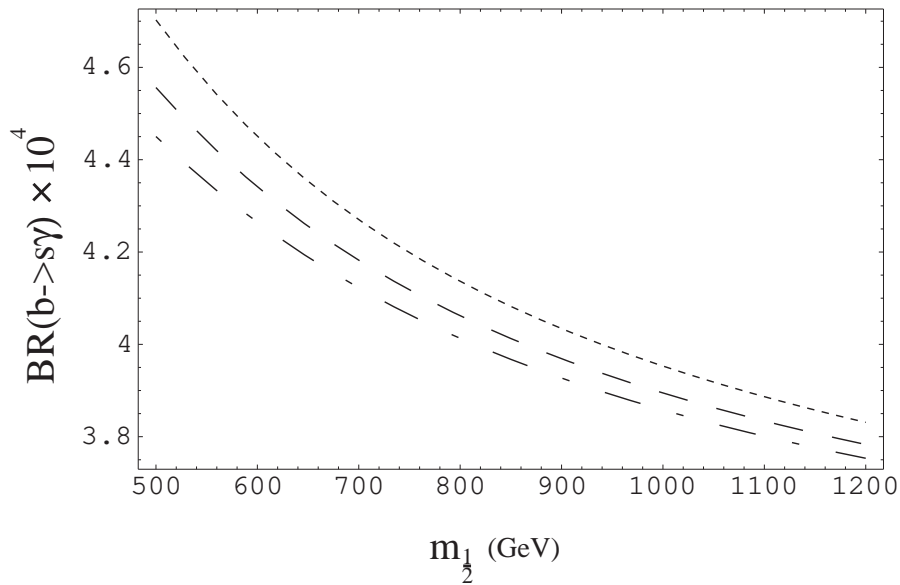

\vspace{1 cm}
\centerline{\DESepsf(bdmneut.epsf width 12 cm) }
\caption{\label{bsgmfig}BR of $b\rightarrow s+\gamma$ is plotted as a
 function of the universal gaugino mass ($m_{1/2}$) for
$\tan\beta=41$. The dotted line is for $\delta_1=$3, $\delta_2=$-0.9
 and $m_0=600$ GeV. The dashed line is for $\delta_1=$2.5, $\delta_2=$-0.9
 and $m_0=700$ GeV. The dash-dotted line is for $\delta_1=$2, $\delta_2=$-0.9
 and $m_0=800$ GeV. }
\end{figure}

The constraint
$\lambda_t=\lambda_b$ at the left--right scale has  further restrictions. The
equality of the  Yukawa coupling makes it impossible to have an experimentally
allowed $m_b$ mass ($m_b=4.25\pm 0.2$) for low and intermediate values of
$\tan\beta$. We find that  $\tan\beta$ $\sim 36-45$ is allowed by the
experimental constraint for reasonable values of $\delta$. The one loop
correction is also included for the evaluation of the $m_b$  and this loop
contribution involves a gluino mediated diagram (involving sbottom) and the
chargino mediated diagram (involving stops). The one loop correction is given by
\cite{rs,wc}:
\begin{eqnarray} \Delta m_b \simeq
-[{2\alpha_{3}\over(3 \pi)} M_{\tilde G}\mu\tan\beta
g({\tilde m}_{b_1}^2,{\tilde m}_{b_2}^2, M_{\tilde G}^2)+
        {\lambda_t^2\over(4\pi)^2}A_t\mu\tan\beta ~g({\tilde m}_{t_1},
    {\tilde m}_{t_2},\mu^2)]
    \lambda_bv_d
\end{eqnarray} where $g(a,b, c)={{(a b Log[{a\over b}]+b c Log[{b\over c}]+
        a c Log[{c\over a}])}\over {(a-b)(b-c)(a-c)}}$ and $M_{\tilde G}$ is the
gluino mass, ${\tilde m}_{b_i}$ are the sbottom  and  ${\tilde m}_{t_i}$ are
the stop masses. The sign of $\mu$ is
needed to be positive (in our convention)
 in order to have correct value of $m_b$. But it has been
also realized \cite {rs,wc,br} that for this choice of the sign of $\mu$,
satisfying the CLEO
bound ($1\times 10^{-4}<{\rm BR}(b\rightarrow s+\gamma)<4.2
\times 10^{-4}$\cite{cleo}) on BR($b\rightarrow s+\gamma$)  requires larger values
of $m_0$ and
$m_{1/2}$. The reason for this is that the chargino diagram adds
constructively to the SM and the Charged Higgs mediated diagrams. In
Fig~\ref{bsgmfig}, we plot the BR$(b \rightarrow s +\gamma)$
as a function of $m_{1/2}$. We have shown
curves for different values of $m_0$ and $\delta_i$. We find that there exists
parameter space for $m_0>600$ and $m_{1/2}>600$ which is allowed by this rare
decay.

\begin{figure}[htb]
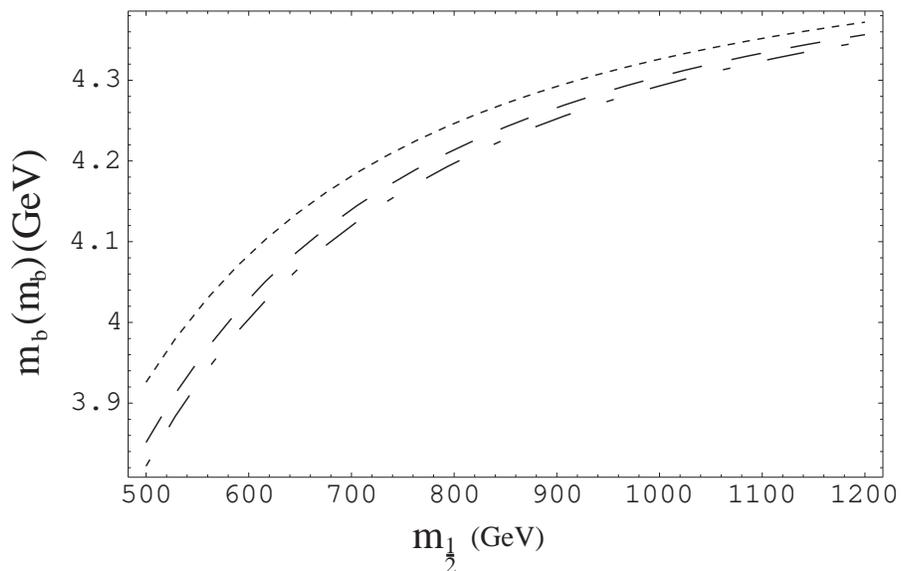

\vspace{1 cm}
\centerline{\DESepsf(bdmneut3.epsf width 12 cm) }
\caption{\label{mbmass}$m_b(m_b)$is plotted as a
 function of the universal gaugino mass ($m_{1/2}$) for
$\tan\beta=41$. The parameters for the curves are same as in fig. 1}
\end{figure}

 In Fig~\ref{mbmass}, we plot the total mass of the $b$--quark ($m_b$),
which includes the one--loop correction,
 as a function of $m_0$. We have drawn  curves for two different values of
 $m_{1/2}$ for $\tan\beta=$41.

\begin{figure}[htb]
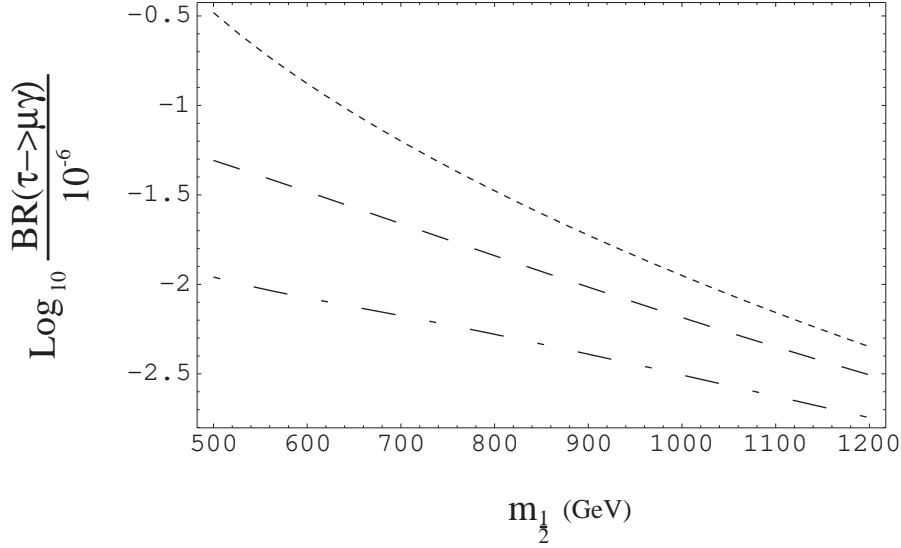

\vspace{1 cm}
\centerline{\DESepsf(bdmneut2.epsf width 12 cm) }
\caption{\label{tmgfig}BR of $\tau\rightarrow \mu+\gamma$ is plotted as a
 function of the universal gaugino mass ($m_{1/2}$) for
$\tan\beta=41$. The parameters for the curves are same as in fig. 1}
\end{figure}

We  now present  our calculations of BR$(\tau\rightarrow\mu+\gamma)$ in the
allowed
 regions of parameter space.
  The experimental upper limit on the  BR$(\tau\rightarrow \mu+\gamma)$ mode
is
$3\times 10^{-6}$  at the 90 $\%$ C.L.\cite{tamu}.  This is expected to improve in the near
future\cite{simone}. In Fig.~\ref{tmgfig}, we plot
$k\equiv Log_{10}{{BR(\tau\rightarrow\mu+\gamma)}\over 10^{-6}}$ as a function of
$m_{1/2}$. We show curves for different values of $m_0$.  We find that,  as
$\delta_1$ increases and $\delta_2$ decreases, the
BR($\tau\rightarrow\mu+\gamma$) gets enhanced (the reason is that the righthanded
and the left--handed stau mass gets lowered). We also find from the figures
 that the BR$(\tau\rightarrow \mu+\gamma)$ is within one or two order of magnitude below the
experimental value. In this model we also have $\mu\rightarrow e+\gamma$ mode, but
the branching ratio is too small (almost 5-6 order of magnitude below the
experimental observation). The reason for this is that the first and second
generation Majorana Yukawa couplings are much smaller than the third generation
ones.  Consequently, the first and second generation
slepton masses are not as suppressed as the third generation slepton
masses.

\begin{figure}[htb]
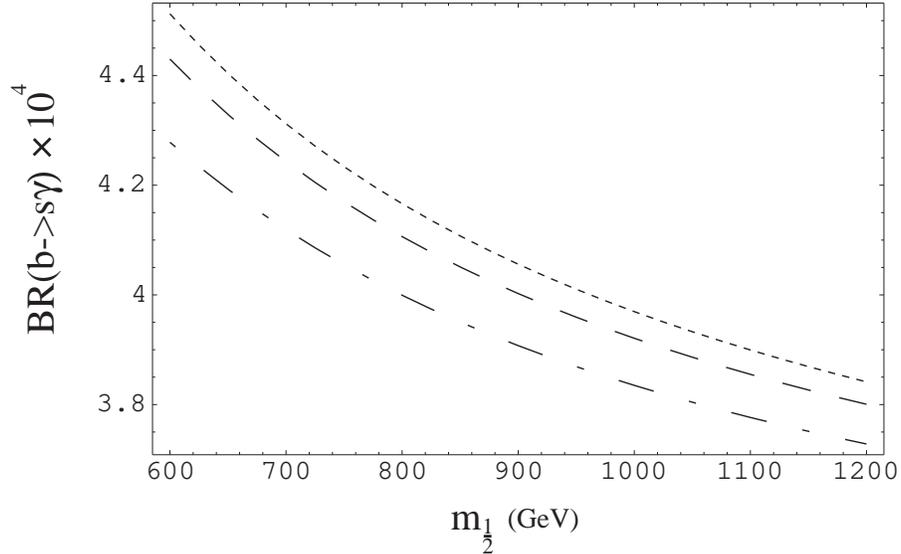

\vspace{1 cm}
\centerline{\DESepsf(bdmneut1A.epsf width 12 cm) }
\caption{\label{bsgmfigA}BR of $b\rightarrow s+\gamma$ is plotted as a
 function of the universal gaugino mass ($m_{1/2}$) for
$\tan\beta=40$. The dotted line is for $\delta_1=$3, $\delta_2=$-0.9,
 $m_0=650$ GeV and $A_0=-550$ GeV. The dashed line is for $\delta_1=$3,
 $\delta_2=$-0.9,
 $m_0=700$ GeV and $A_0=-600$ GeV. The dash-dotted line is for $\delta_1=$2.5,
 $\delta_2=$-0.9,
$m_0=800$ GeV and $A_0=-650$ GeV.}
\end{figure}

\begin{figure}[htb]
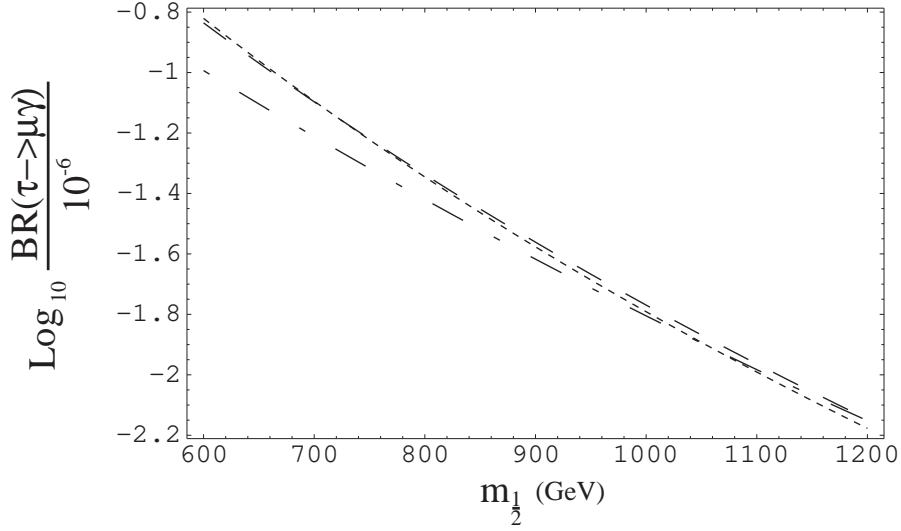

\vspace{1 cm}
\centerline{\DESepsf(bdmneut2A.epsf width 12 cm) }
\caption{\label{tmgfigA}BR of $\tau\rightarrow \mu+\gamma$ is plotted as a
 function of the universal gaugino mass ($m_{1/2}$).
 The parameters for the curves are same as in fig. 4}
\end{figure}

So far we have given our results for universal trilinear term $A=0$. If we
assume $A$ to be non--zero, we can lower the $m_b$ mass correction (since the
chargino contribution increases the mass correction). The allowed
parameter space of this model would increase. However one should be careful
about the lighter stau mass. The change in $A_t$ at the weak scale is much less
than the change in $A_{\tau}$  for a non--zero
$A_0$.  Consequently the lighter stau mass--squared (which is already light in this
model) can turn negative. However one also can choose $A$ judiciously, so that
the stau mass is  lighter (but not lighter than the lightest neutralino) and as a result the
BR$(\tau\rightarrow\mu+\gamma)$ will increase. We show examples of such
case in Figs~\ref{bsgmfigA},~\ref{tmgfigA}. In these figures we can see that the
BR($b\rightarrow s+\gamma$) does not change at all compared to the
Fig~\ref{bsgmfig}, but the
BR$(\tau\rightarrow \mu+\gamma)$ changes quite a bit.

\begin{figure}[htb]
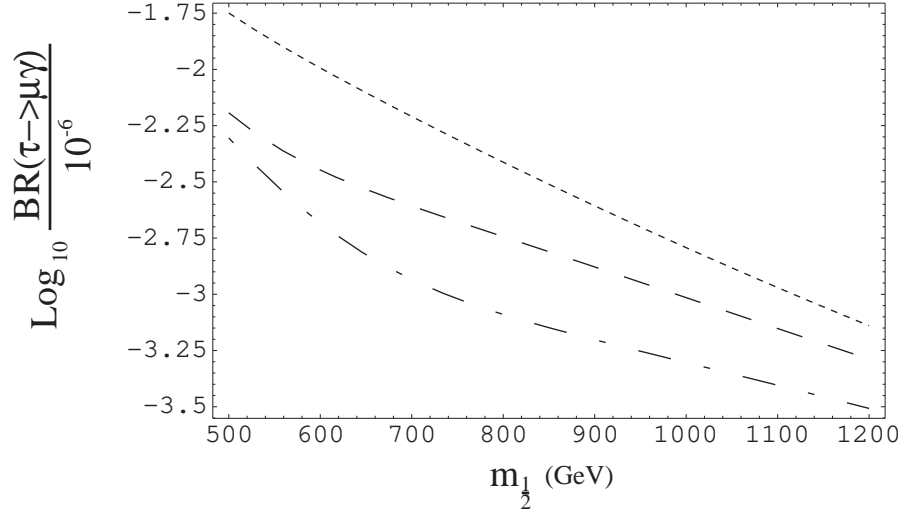

\vspace{1 cm}
\centerline{\DESepsf(bdmneut2nr.epsf width 12 cm) }
\caption{\label{tmgfignr}BR of $\tau\rightarrow \mu+\gamma$ is plotted as a
 function of the universal gaugino mass ($m_{1/2}$) for
$\tan\beta=41$ in the case when doubly charged fields get decoupled at the left
right scale. The dotted line is for $\delta_1=$3, $\delta_2=$-0.9
 and $m_0=600$ GeV. The dashed line is for $\delta_1=$2.5, $\delta_2=$-0.9
 and $m_0=700$ GeV. The dash-dotted line is for $\delta_1=$2, $\delta_2=$-0.9
 and $m_0=800$ GeV.}
\end{figure}

In the Fig.~\ref{tmgfignr}, we plot the BR$(\tau\rightarrow\mu+\gamma)$ in the
case when the doubly charged Higgs fields get decoupled at the left--right scale.
This is the situation that would arise in the presence of $\eta, \overline{\eta}$ fields,
as discussed earlier.
The slepton mass matrix will have less flavor violation in this case and the
$Br(\tau\rightarrow\mu+\gamma)$ is smaller than the previous cases. The
$Br(b\rightarrow s+\gamma)$ remains unchanged since the squark sector does not couple
to the doubly charged fields.

In conclusion, we have shown that in the left--right models for neutrino masses,
the up--down unification allows a fit to both the solar and atmospheric neutrino
data with a hierarchical set of Majorana Yukawa couplings of the leptons. In the
resulting parameter space, one obtains a prediction for the rare decay
$\tau\rightarrow
\mu+\gamma$ which is within the accessible range of the planned experiments.

The work of KSB is supported by funds from the Oklahoma State University.
RNM is supported by the National Science Foundation grant No.
PHY-9802551.

\end{document}